# Polarization and Space Charge Limited Current in III-Nitride Heterostructure Nanowires

Michael A. Mastro, Hong-Youl Kim, Jaehui Ahn, Blake Simpkins, Pehr Pehrsson, Jihyun Kim, Jennifer K. Hite, and Charles R. Eddy, Jr.

M. A. Mastro, Blake Simpkins, Pehr Pehrsson, Jennifer K. Hite, Charles R. Eddy, Jr: U. S. Naval Research Lab
Hong-Youl Kim, Jaehui Ahn, Jihyun Kim: Department of Chemical and Biological Engineering, Korea University

*Abstract*— An undoped AlGaN/GaN nanowire demonstrated p-type conductivity based solely on the formation of hole carriers in response to the negative polarization field at the (000-1) AlGaN/GaN facet. A transistor based on this nanowire displayed a low-voltage transition from Ohmic to space charge limited conduction. A numerical simulation showed that a highly asymmetric strain exists across the triangular cross-section, which creates a doublet peak in the piezoelectric induced polarization sheet charge at the (000-1) facet. Additionally, there is a strong interplay between the charge at the (000-1) AlGaN/GaN interface with depletion from the three surfaces as well as an interaction with the opposing polarization fields at two semi-polar {-110-1} facets. The charge distribution and resultant conduction regime is highly interdependent on configuration of the multi-layer structure, and is not amenable to an analytical model.

## I. INTRODUCTION

The phenomena of space charge limited current (SCLC) was noted half a century ago in several of the first relatively-pure semiconductor crystals that were fabricated into two-terminal devices. [1] The low concentration of carriers created an extended spatial curvature in the electrostatic potential, $\psi$.

By Poisson's equation, $\frac{d^2\psi}{dx^2} = \frac{-d\xi}{dx} = -\frac{\rho}{\varepsilon}$, a gradient in electric field, $\xi$, will manifest as an additional charge $\rho$ if the thermally generated level of carriers is insufficient. Under Ohmic conduction conditions, the drift current density $J_{Drift} = \mu\rho\xi$ is determined by the mobility, $\mu$, charge, and applied voltage over a channel of length L, assuming $\xi \approx \frac{V}{L}$. In a region depleted of carriers, the curvature of the potential generates additional carriers by $\rho = \varepsilon\frac{d\xi}{dx}$ and a corresponding increase in the current density $J = \mu\varepsilon\frac{d\xi}{dx}\xi$. [2] When this region extends across the entire current path, e.g., source to drain or cathode to anode, the device is controlled via this SCLC. Surface depletion in thin-film structures is normally limited to one surface. In contrast, the equivalent nanowire structure often suffers from pronounced surface depletion from its inherently large surface to volume ratio. This manifests as SCLC in lightly doped nanowires for even a moderate bias voltage.

An analytical model of SCLC in a cylindrical homogeneous GaN nanowire was recently presented, [3] however, these equations are not adequate for III-nitride heterostructure nanowires because GaN nanowires tend not to be cylindrical. For example, the standard approach to produce GaN nanowires via a vapor-liquid-solid mechanism forms a nanowire along the a-plane direction with an isosceles triangle cross section with two identical {-110-1} facets and a third {0001} facet. The large polarization constants in the III-nitride material system yield large and dissimilar polarization-induced hole or electron accumulation along the interfaces at one or multiple facets. This article presents two and three-dimensional numerical simulations that confirm the experimental findings that AlGaN/GaN shell/core nanowires can be designed to operate in an Ohmic or a SCLC regime.

## II. Experimental

NW transistor fabrication: A 0.05 M nickel nitrate solution was repeatedly dripped onto a r-plane sapphire substrate and blown dry in $N_2$ then loaded into a vertical impinging flow, metal organic chemical vapor deposition reactor. The GaN nanowire core was grown at a temperature of 850°C. The growth kinetics of a GaN nanowire at the higher growth temperature favored formation of nanowires with a (000-1) N-face facet as apposed to the (0001) Ga-face facet. [4, 5] In vapor-liquid-solid mechanism, the metal seed acts as a catalyst for one-dimensional vertical growth, and the seed diameter determines the altitude of the GaN. Despite the rapid vertical growth rate, a small flux of reactants will deposit on the sidewalls creating a slight taper along the length of the nanowire. On average, the GaN core has a base to vertex altitude of 50nm. The $Al_{0.25}Ga_{0.75}N$ shell was grown at a higher temperature of 950°C to encourage in-plane horizontal growth. Under the $Al_{0.25}Ga_{0.75}N$ growth conditions, the energetics of the (000-1) N-face surface readily incorporate atoms from the growth environment resulting in a slightly higher growth rate and thickness (20 nm) compared to the thickness of the semipolar surfaces (10 nm). The thickness and composition of the AlGaN layer was selected to ensure pseudomorphic growth. [12]

The III-nitride nanowires were dispersed in an iso-propyl alcohol solution by sonication, and subsequently dispersed onto a 300nm $SiO_2$ / Si (thickness: 525 um; resistivity: 5 ohm-cm) wafer. The $SiO_2$/Si wafer was pre-patterned with finger-shaped Ti/Au electrodes (20 nm/80 nm thickness). Electron-beam lithography was used to form the Ni/Au contacts (20nm/100nm; annealed at 400 C for 30 s) to connect the source and drain regions of the nanowire to large area contact pads. A Ni/Au layer was deposited on the backside of the conductive wafer to provide a gate electrode for the NW-FET. The source-to-drain distance was 500nm. The simulated structure in figure 2d, e, f and 3 is similar to the experimental structure.

## III. Results and Discussion

Current-voltage characteristics of an undoped AlGaN/GaN p-type nanowire transistor are presented in Figure 1. The increase in conduction for decreasing gate bias confirms the nanowires are p-type in nature. The native conductivity of undoped GaN is n-type from deep point and extended defects [6]. The p-type conductivity is therefore only attributable to the negative fixed polarization charge at the (000-1) AlGaN/GaN interface, which induces the accumulation of positively charged holes. [4] The variation in strain at this abrupt heterojunction creates a piezoelectric polarization field, $P_{pz}$, and variation in composition at this interface creates a discontinuity in the spontaneous polarization field, $P_{sp}$. The total polarization ($P=P_{sp}+P_{pz}$) incorporates into Poisson's equation as $\nabla \cdot D = \nabla \cdot (\varepsilon \xi + P) = \rho$, where D is the displacement vector and $\varepsilon$ is the permittivity. Expressing with the electrostatic potential gives $\nabla \cdot [\varepsilon(-\nabla \psi)] + \nabla \cdot P = \rho$. When $\nabla \varepsilon \to 0$, then

$$\nabla^2 \psi = -\frac{\rho}{\varepsilon} + \frac{1}{\varepsilon}[\nabla \cdot P].$$

Assuming that the polarization occurring at hetero-interface results in a fixed polarization charge, $\rho^{Pol}$, gives

$$\nabla^2 \psi = -\frac{\rho}{\varepsilon} - \frac{\rho^{Pol}}{\varepsilon}.$$

The charge, ρ, in the semiconductor is primarily composed of hole carriers, p, electron carriers, n, ionized donors, $N_D^+$, and ionized acceptors $N_A^-$; however, other effects are often incorporated including donor and acceptor traps. [7] Substituting these factors gives the functional form of Poisson's equation as
$$\nabla^2 \psi = -\frac{q}{\varepsilon}[N_D^+ - N_A^- + p - n] - \frac{\rho^{Pol}}{\varepsilon},$$
which describes the formation of charged carriers in semiconductor in response to the potential field.

In general, the electron current density is composed of drift and diffusion terms, $J = \mu \rho \xi \pm C_D \frac{d\rho}{dx}$, where $\mu$ is the mobility, $C_D$ is the diffusion constant, and the charge density is $\rho = qn$ for n electrons or $\rho = qp$ for p holes. In typical bulk and thin film semiconductor unipolar devices, the electrostatic-potential $\psi$ drops linearly from the source to drain. This is equivalent to stating that the electric field is constant, $\xi = -\nabla \psi \approx V/L$. Implicit in these two previous statements is that a sufficient number of dopant atoms are available to thermally generate mobile carriers to maintain current continuity. Ohmic conduction is typically well described by the linear relation, $J \approx \mu \rho \xi \approx \mu \rho \frac{V}{L}$, while the carrier velocity is far from saturation and the applied field is far below the critical breakdown-voltage. This first order dependence on voltage is readily revealed in a data set with a slope of approximately one in a log current – log voltage plot. Figure 1b shows that Ohmic conduction is dominant at low voltage but transitions to a quadratic dependence, i.e., slope of approximately two, corresponding to an SCLC regime, at higher voltage.

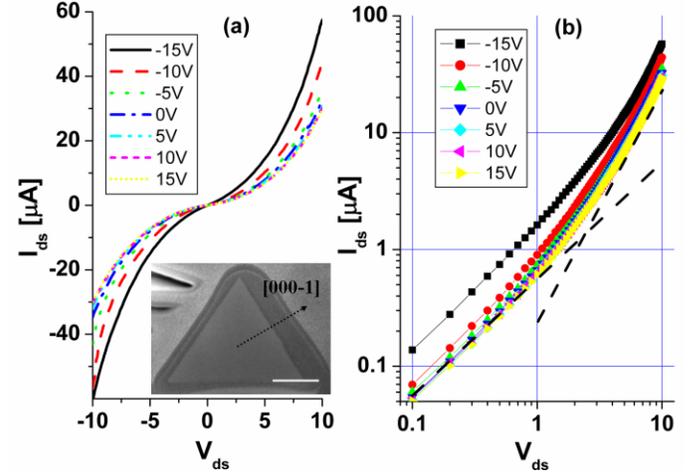

Figure. 1. (a) The increase in drain-source current-voltage of a p-type AlGaN/GaN nanowire for decreasing gate bias confirms transport is via hole conduction. Inset: Cross-sectional electron micrograph of the nanowire. Scale bar 30nm. (b) A log current - log voltage plot reveals a transition from a slope of approximately 1 to a slope of approximately 2. Ohmic conduction is dominant at low drain-source voltage while space charge limited current is dominant at higher voltage. The dotted straight lines are used as a guide to delineate these two conduction regimes, linear $I \propto V$, and quadratic $I \propto V^2$. The intercept of low voltage and high voltage asymptotic lines defines the threshold voltage for transition between the conduction regimes that increases from 2.2V to 3.9V over gate bias range (15V to -15V).

In this SCLC regime, an excess of charge carriers are injected from the contacts far beyond the available number of ionized carriers. This situation can be described by a simplified form of Poisson's equation, $\frac{d\xi}{dx} = \frac{\rho}{\varepsilon}$, where the curved, i.e., non-constant, electric field determines the charge density. Substituting the charge from Poisson's equation into the current density drift equation gives, $\xi d\xi = \frac{J}{\mu\varepsilon} dx$. Integrating gives the relation, $\frac{1}{2}\xi^2 = \frac{Jx}{\mu\varepsilon} + const$. Using the boundary conditions that the source is at ground, $\psi(0) = 0$, and drain is at fixed voltage, $\psi(L) = V$, along with an additional integration gives $\psi = V = \sqrt{\frac{8J}{9\mu\varepsilon}} L^{3/2}$. Rearranging shows the well-known quadratic voltage dependence for SCLC in a bulk semiconductor, $J = \frac{9\mu\varepsilon}{8} \frac{V^2}{L^3}$. At higher-voltage, the current data asymptotically approaches a slope of approximately 2. It is well understood in bulk crystals that an increase in the slope (i.e., slope above two) reveals the energetic distribution of the traps and the intercept is a measure of the ratio of trapped to free conduction states. Nevertheless, bulk Ohmic and SCLC equations are not directly applicable to nanowires, and application of bulk equations to nanowires may yield misleading conclusions. Additionally, the interplay of charge and strain in III-nitride triangular heterojuncton nanowires further complicates the analysis.

Simpkins found strong diameter dependence in Ohmic conduction in GaN nanowires. [8] Therein, Simpkins developed a model where the Ohmic conduction in homogenous GaN is based on a conductive core with no-contribution from a surface-depleted shell. Talin derived an analytical SCLC equation, $J = \zeta(R/L) \mu\varepsilon \frac{V^2}{L^3}$, that affixed a (circular) geometrical $\zeta(R/L)$ scaling factor to account for surface depletion that reduces to the bulk 9/8 value for large diameter nanowires. [3] Additionally, Talin predicted that the Ohmic to SCLC cross-over voltage, $V_c = \frac{1}{\zeta} \frac{qnL^2}{\varepsilon} \approx (R/L)^2 V_c^{bulk}$, is significantly reduced for high-aspect nanowires. In other words, a nanowire will display at moderate voltages a quadratic SCLC response while the equivalent thin film device will display Ohmic conduction. At an extreme scaling of aspect ratio, a thin nanowire would only operate in SCLC.

Simpkins recently measured SCLC in n-type AlGaN/GaN nanowires grown under lower temperature to invert the cross-sectional polarity as compared to the nanowire presented in this article. [9] The geometry consisted of the (0001) facet where the positive polarization charge at the AlGaN/GaN interface accumulated electrons on the GaN side of the interface as a two dimensional electron gas.[4] The approximately 100nm facet width allowed Simpkins to describe the SCLC as a sheet of conduction similar to a thin film device.

The nano-sized triangular cross-sectional geometry, typical for a GaN-based nanowire grown from a metal catalyst, introduces asymmetric strain and surface depletion as well as interaction with the opposing polarization charge at the two semi-polar planes. Figure 2 displays the electron and hole density, valence band edge, and x-x strain tensor component for three different AlGaN shell thicknesses with a constant GaN core thickness based on a coupled numerical calculation.

In the calculation a background n-type conductivity of $1 \times 10^{15}$ cm$^{-3}$ was assumed from unintentional point defects that unavoidably form in GaN. [10] Similarly, n-type surface traps were added to the AlGaN surface. Any surface of a semiconductor inherently possesses broken bonds that reconstruct to minimize the free energy of the local system. The energetics in (Al)GaN encourages formation of n-type surface states. [11] As the overall cross-sectional size is reduced, the surface states deplete the outer edge of the hole gas at the (000-1) interface. This first affects the edges of the sheet (adjacent corners of the triangle) to create a Gaussian-like shape but eventually depletes the entire sheet charge as the thickness of the shell is further reduced.

In [4], Mastro developed an analytical model to calculate the amount of polarization and associated fixed polarization charge at the {1-101} facets that is opposite to the charge on the corresponding {0001} facet. This is not mobile charge and the amount of polarization-induced charge (free electrons or holes) that is present depends on the Fermi level in the bulk material and barrier height. Although an analytical model can show general trends, it is necessary to undertake a numerical solution to account for the non-uniform strain and associated piezoelectric charge, and its influence on the Fermi level across the wire.

The positive polarization charge along the two semi-polar {1-101} facets is smaller in magnitude than the negative polarization charge at the (000-1) facet. The positive polarization charge along the two semi-polar {1-101} facets is not large enough to accumulate electrons but it does aid in depleting the edges of the hole gas at the (000-1) interface. The moderate shell thickness (figure 2d) shows the hole sheet charge is further reduced by surface depletion and the opposing polarization charge at the semi-polar facet offsets the change in local Fermi level.

For a thin shell, as displayed in figure 2g, a small yet significant concentration of electrons can appear at the apex of the two semi-polar {1-101} facets. Here, the surface states pull up the Fermi level to deplete the (000-1) hole gas and allow the dual contribution of positive polarization charge along the two semi-polar {1-101} facets to accumulate electrons at the apex.

The two-dimensional graphical representation of the valence band is displayed in figure 2b, e, and h, with an energy scale relative to the Fermi level at 0 V. Inspection of figure 2b and e shows that the valence band is approximately 0 V near the mid-point of the GaN side of the (000-1) interface. In fact, the valence band forms a potential well a few meV below the Fermi level, which is characteristic of a two-dimensional hole gas. Inspection of the bottom vertex of figure 2h shows that the valence band edge is pulled 3.4 eV below the Fermi level. For brevity, the conduction band edge is not displayed but since the band gap of GaN is 3.4eV then it follows that the conduction band edge is near the Fermi level at 0 V. Consequently, there is an accumulation of electrons at the bottom vertex as seen in figure 2g.

Conceptually, a large AlGaN/GaN microwire with a facet width of several microns would display strain, polarization and charge similar to its equivalent thin film structure. A nano-size effect is that significant non-uniform strain will develop at the vertexes of the triangle. It should be noted that $\varepsilon_{xx}$ in fig 2c, g, and i is one component of the strain tensor. Rotating the wire within the coordinate domain by 120° would produce a similar $\varepsilon_{xx}$ plot. That is, if slight thickness variation is neglected, each face of the nanowire experiences similar strain.

In general, when AlGaN is deposited pseudomorphically on GaN, the lattice mismatch will create tensile strain in the AlGaN film. [12] If both layers are thin, both layers will contribute to strain accommodation by curving the structure like a bowl. In a simplified view, the strain state in the triangular AlGaN/GaN wire is created by mating three strained systems. Thus, there is a peak in tensile strain near the apex and resultant increase in piezoelectric-polarization charge. For a small GaN core, this strain enhancement is sufficient to create two spikes in the holes gas near the (000-1) apexes as seen in figure 2a and b. A thicker shell (as in figure 2c) occupies a majority of the material volume and the smaller interior core will distort to match the shell.

In contrast, a thin 10nm AlGaN shell (as in figure 2i) occupies a smaller fraction of material volume and is highly strained throughout its volume. Similarly, additional simulations show that large GaN core nanowires have reduced strain asymmetry and thus may not exhibit this localized carrier doublet peak (not shown).

These trends in carrier distribution will have marked impact on observed conduction mechanisms. Below, we calculate band profiles for coaxial heterostructures NWs with strain states similar to figure 2d, e, and f (with a moderate 20nm shell) under various charge injection conditions and discuss the impact of nanostructure on conduction mechanism. In general, the trends in simulated current and Ohmic to SCLC voltage threshold match the experimental results. A voltage offset arises that is attributed to small differences in the simulation from the experiment, particularly, a back-side gate displaced spatially from the nanowire.

The transition from Ohmic to SCLC is understood by examining figure 3, which displays electrostatic properties along the axial direction of a gated AlGaN/GaN nanowire transistor. The data is plotted along a line in the center of the hole gas, orthogonal to the cross-sections of figures 2d,e, and f.

In figure 3i, an unbiased Schottky gate has partially depleted the hole gas from approximately 100 to 400nm to create a p/p⁻/p structure. The left-most column shows zero bias from source to drain, and thus is symmetric in the valence band energy. In figure 3a, the virtual cathode is located at the center of the valence band, which corresponds to the zero point crossing of the electric field ($\xi$ in dashed line in figure 3e). The electric field describes the curvature of the electrostatic potential, thus the virtual cathode can equivalently be defined as the minimum of the electrostatic potential ($\psi$ in solid line in figure 3e).

The device is under Ohmic conditions at a small (0.1V) source-drain voltage visualized in the second column. From figure 3a to figure 3d, the virtual cathode has shifted closer to the source cathode. The energy barrier at the virtual cathode in figure 3b is similar to the top of the barrier models used to describe short channel field effect transistors. [13]

The third column shows that sufficient source-drain voltage has been applied to shift the virtual cathode completely to the source cathode. The basis of most analytical models of SCLC is the virtual cathode approximation, which assumes diffusion transport on the anode side and variation of the quasi Fermi-level on the cathode side of the virtual cathode can be ignored. [14] As discussed above, the semiconductor responds to a curvature of the potential by generating additional charge (or carriers) according to $\rho = \varepsilon \frac{d\xi}{dx}$. A further increase in drain-voltage in the fourth column shows an even larger increase in carriers and a corresponding increase in current density via $J = \mu \varepsilon \frac{d\xi}{dx} \xi$.

The quadratic $J \propto V^2$ dependence of SCLC transforms to a $J \propto V$ when the carriers attain velocity saturation. The SCLC effect is often present but not apparent in semiconductors with low velocity saturation voltage such as GaAs (0.72 kV/cm) as compared to GaN with its higher velocity saturation voltage (1.4 kV/cm). This velocity-saturated SCLC is commonly observed as the punch-through mechanism in bipolar transistors.

One feature of this article is that the triangular surface of a III-nitride nanowire will enhance carrier depletion and promote conditions for SCLC. An additional feature that encourages SCLC in GaN in its thin film, bulk, or nanowire form is its higher breakdown voltage (5 MV/cm), which allows the conditions of SCLC to be reached under voltage conditions that would correspond to avalanche or other breakdown mechanism in GaAs (0.4 MV/cm). Recent models suggests that high-voltage AlGaN/GaN thin-film transistors are hampered below threshold by a nonlinear resistance associated with SCLC in the gate-source region, [15] and have threshold controlled by SCLC at the drain-edge of the gate. [16]

Additionally, a high-aspect low-doped GaN nanowire transistor naturally operates as a static induction transistor without the deleterious etching steps required for its thin film equivalent. It is reasonable to expect that control or elimination of SCLC will govern the design of all III-nitride nanowire transistors.

IV. CONCLUSION

An undoped AlGaN/GaN nanowire demonstrated p-type conductivity based solely on the formation of hole carriers in response to the negative polarization field at the (000-1) AlGaN/GaN facet. A transistor based on this nanowire displayed a low-voltage transition from Ohmic conduction to SCLC. A numerical simulation showed that a highly peaked charge distribution can form within the wire from the interaction of asymmetric strain, surface depletion, and opposing polarization fields at the two {-110-1} facets. The complexity of the strain and charge distribution implies that the conductive behavior of the III-nitride heterostructure nanowire can only be understood via a coupled numerical simulation

REFERENCES

[1] R.W. Smith, A. Rose, Phys. Rev. 97, 1531 (1955)
[2] G.T. Wright, Solid-State Electronics, 2, 165

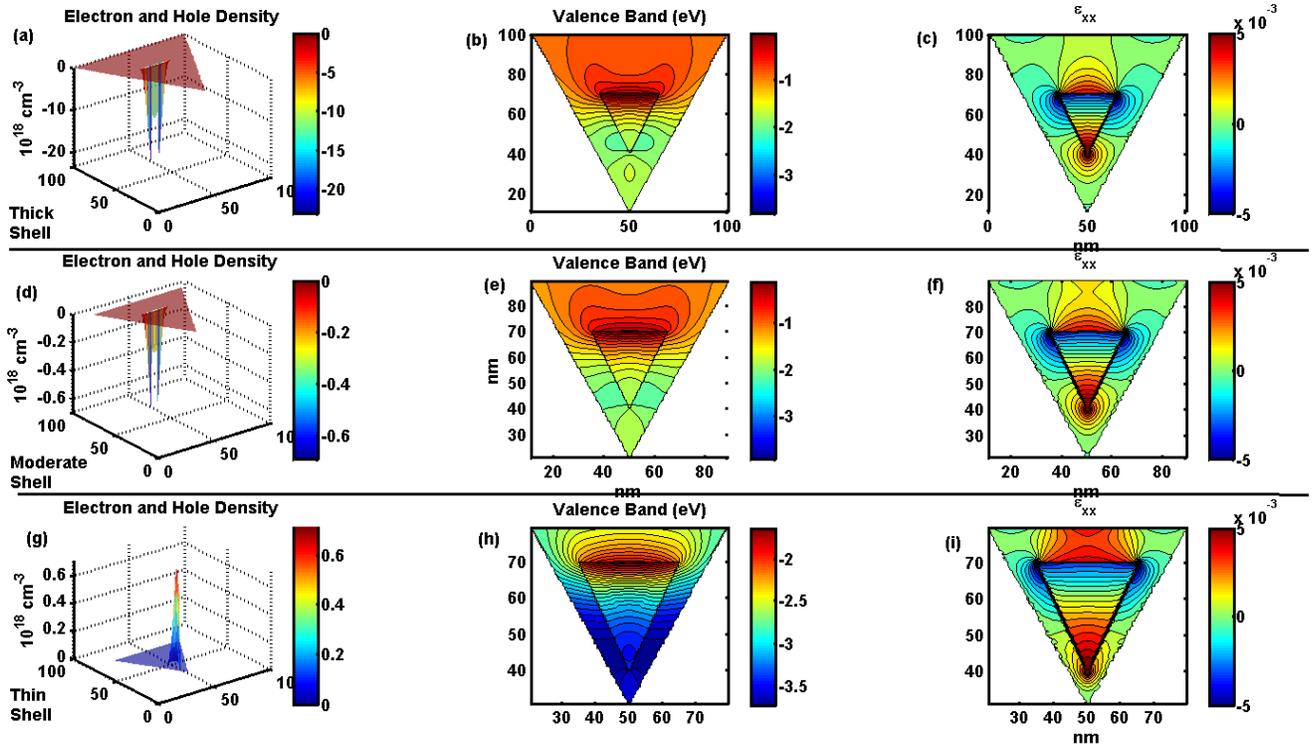

Figure 2. Electron and hole density, valence band edge, and x-x strain tensor component for three different (000-1) AlGaN shell thicknesses (top 30nm; middle 20nm, bottom 10nm) with a constant GaN core thickness. Note that the x-axis and y-axis maximum is 80, 90, or 100 nm to match the size of the nanowire. The (000-1) facet is at the top of each graph while two semi-polar (1-101) and (-1101) facets are inclined towards the bottom of the graph. In (a,d, and g) the scale is such that negative concentration implies holes and positive concentration implies electrons. A decrease in shell thickness allows the surface states to further deplete the hole gas at the (000-1) interface. Note that the z-axis scale in (a) is approximately an order of magnitude larger than (d) and (g). The asymmetry in the strain fields creates an asymmetry in the piezoelectric-polarization induced holes that are observable as doublet-peaks in (a) and (d).

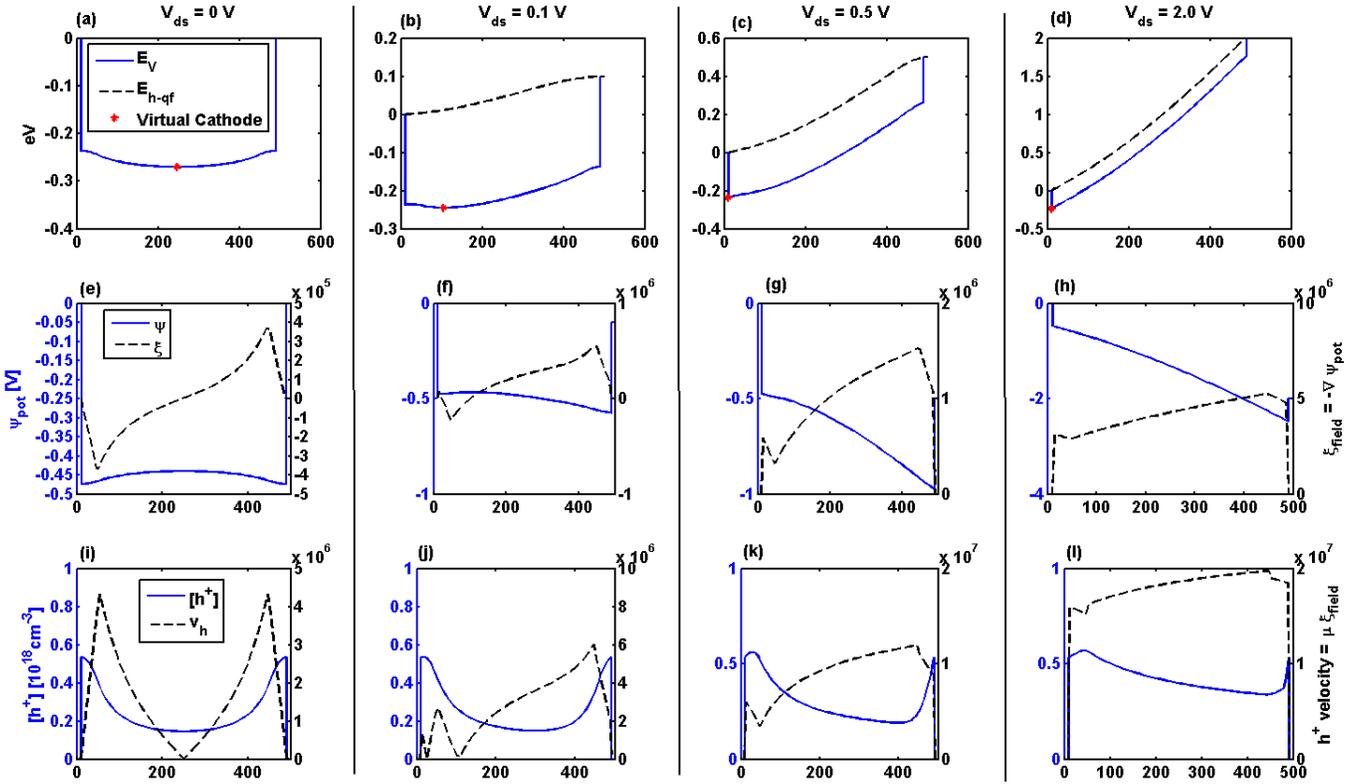

Figure 3. Electrostatic profiles along to the (000-1) AlGaN/GaN interface (approximately x=50nm, y=70nm in figure 2d). The source (cathode) and drain (anode) are located at 5nm and 495nm, respectively. Each column displays the device at increasing source-drain bias. (a, b, c and d) Valence band, quasi-Fermi level and virtual cathode point. (e, f, g, and h) Electrostatic potential, $\psi$, and electric field, $\xi = \dfrac{-d\psi}{dx}$. (i, j, k, and l) Hole concentration and drift velocity. The unbiased gate electrode depletes the carriers along the length of the nanowire (from approximately 50 to 450 nm) forming an pp⁻p structure. This creates an electrostatic potential gradient, i.e., electric field, that peaks near either end of the nanowire. The holes drift in response to the electric field that has a zero point crossing (i.e., sign change) at the virtual cathode. An equilibrium state is attained by hole diffusion from a high to a low carrier concentration, opposite to the electric field. A small 0.1V bias on the drain shifts the potential profile across the wire with the virtual cathode now (at the electrostatic potential minimum or electric field zero point) located closer to the source. At $V_{ds}$ = 0.1V, holes on the source side of the virtual cathode will drift towards the source. Further increasing the drain bias causes the potential to continuously decrease from the source to the drain, and carrier drift only occurs from the source towards the drain.